\documentclass[aps,pre,showpacs,twocolumn,floats]{revtex4}
\usepackage{amssymb}

\usepackage{amsmath}
\usepackage{graphicx,psfig}
\usepackage{dcolumn}
\usepackage{bm}

\input{tcilatex}

\begin{document}

\title{Collective electromagnetic relaxation in crystals of molecular magnets}
\author{Collin L. Joseph, Carlos Calero, and Eugene M. Chudnovsky}
\affiliation{Department of Physics and Astronomy, Lehman College,
City University of New York \\ \mbox{250 Bedford Park Boulevard
West, Bronx, New York 10468-1589, U.S.A.}}
\date{\today}

\begin{abstract}
We study the magnetization reversal and electromagnetic radiation
due to collective Landau-Zener relaxation in a crystal of
molecular magnets. Analytical and numerical solutions for the time
dependence of the relaxation process are obtained. The power of
the radiation and the total emitted energy are computed as
functions of the crystal parameters and the field sweep rate.
\end{abstract}

\maketitle
\section{Introduction}

\begin{figure}
\unitlength1cm
\begin{picture}(11,6.5)
\centerline{\psfig{file=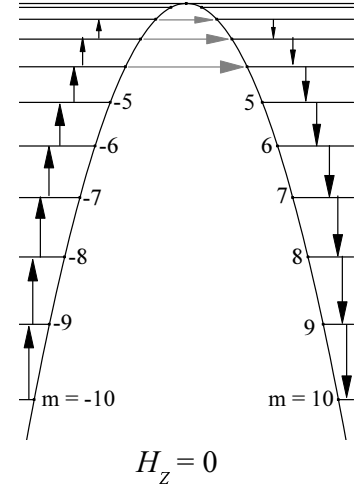,width=4.5cm}}
\end{picture}
\caption{\label{fig:levels} Approximate energy levels of a spin-10
molecule in a zero magnetic field. The tunnel splitting of the
degenerate levels is not shown. Arrows show the relaxation path
from $m=-10$ to $m=10$ through thermally assisted quantum
tunneling. }
\end{figure}

Paramagnetic crystals of high-spin molecular magnets, like
Mn$_{12}$, Fe$_8$ and others, exhibit unusual magnetic properties
related to the macroscopic time of the transition between spin-up
and spin-down states of individual magnetic molecules
\cite{Sessoli}. The latter is due to the high magnetic anisotropy
and a large value of spin, $S \gg 1$. For, e.g., a biaxial
molecule (Fe$_8$ of $S=10$), in the magnetic field, ${\bf H}$,
parallel to the anisotropy axis $Z$, the spin Hamiltonian is
\begin{equation}\label{biaxial}
{\cal{H}} = -DS_z^2 + AS_x^2 -g{\mu}_B H_z S_z\,,
\end{equation}
where $g$ is the gyromagnetic factor, ${\mu}_B$ is the Bohr
magneton, and $D > A > 0$. For small $A$, the approximate energy
states of ${\cal{H}}$ are the eigenstates of $S_z$: $S_z | m
\rangle = m |m \rangle$. At $H_z = kD/g\mu_B$, with $k = 0, \pm 1,
\pm 2, ...$, the levels $m < 0$ and $m'$ satisfying $m + m' = -k$
come to resonance. For the even $S$, the tunnel splitting of the
resonant levels, $\Delta_m$, appears (for even $k$) in the
$[(m'-m)/2]$-th order of the perturbation theory on $A$:
$\Delta_{(m=-S)} \propto (A/D)^{(m'-m)/2}$. At, e.g., $k=0$ (see
Fig.\ \ref{fig:levels}), $\Delta_{(m=-S)} \propto (A/D)^{S}$ and,
thus, at $S=10$, the probability of the transition between spin-up
and spin-down states is low. Consequently, at low temperature, the
crystal can be prepared in a state with inverse population of the
spin energy levels, e.g., magnetized against the direction of the
magnetic field. This allows one to observe, in a macroscopic
experiment, such quantum effects as resonant spin tunneling
\cite{Friedman,confirmation}, spin Berry phase \cite{WS},
crossover between quantum tunneling and thermal activation
\cite{CG,Kent,Sarachik}, and quantum selection rules in the
absorption of electromagnetic radiation \cite{Wernsdorfer}.

\begin{figure}
\unitlength1cm
\begin{picture}(11,6)
\centerline{\psfig{file=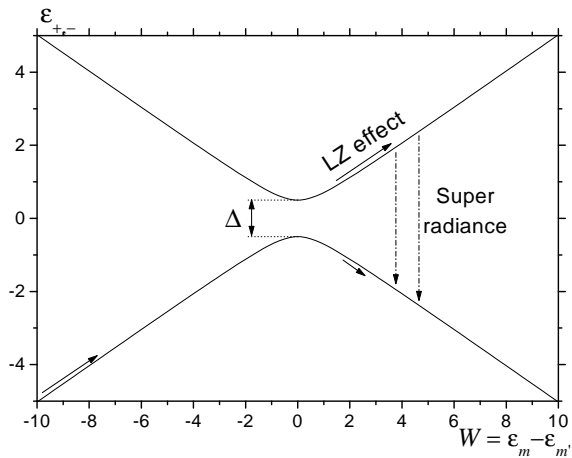,angle=-90,width=9cm}}
\end{picture}
\caption{\label{fig:spl} A pair of tunnel-split levels vs. energy
bias $W$. The Landau-Zener (LZ) transition is followed by the
emission of the coherent light via superradiance. }
\end{figure}

Recently, it has been suggested \cite{CG-SR,Tejada1,Henner} that a
crystal of molecular nanomagnets can be a source of coherent
electromagnetic radiation in the millimeter wavelength range,
highly desirable for applications \cite{Mittleman}. Some
experimental evidence of this effect has been obtained
\cite{Tejada1,Tejada2,Leuven}. The effect is related to Dicke
superradiance \cite{Dicke}. Normally, atoms or molecules of a gas,
initially prepared in the excited energy state, decay
independently by spontaneous emission of light. The power of the
radiation obeys the law ~$P \propto N\exp(-t/{\tau})$~ where ~$N$~
is the total number of atoms and ~$\tau$~ is the lifetime of the
excited state. Dicke argued that $N$ atoms confined within a
volume of size ~$d$~, which is small compared to the wavelength of
the radiation ~$\lambda$, cannot radiate independently from each
other. At ~$d < {\lambda}$~ a spontaneous phase locking of the
atomic dipoles takes place, that results in the coherent radiation
burst of power ~$P_{SR} \propto N^2$~, emitted within a time of
order  ~${\tau}_{SR} \, {\sim} \, {\tau}/N$. This phenomenon,
called superfluorescence, has been widely observed in gases. It
can occur in any system of identical quantum objects if the system
is not very large compared to the wavelength of the radiation
\cite{CG-condmat}. For crystals of molecular magnets this is true
for both, the transitions between the tunnel-split levels, Fig.\
\ref{fig:spl}, and the transitions between the adjacent $|m
\rangle$ levels, Fig.\ \ref{fig:levels}.

In a typical experiment, one magnetizes the crystal and then
sweeps the field in the opposite direction. In this paper we will
be concerned with the situation when the electromagnetic
transitions occur between tunnel-split levels, Fig.\
\ref{fig:spl}. These can be, e.g., transitions between $m=-10$ and
$m=10$ levels shown in Fig.\ \ref{fig:levels}. The electromagnetic
relaxation of that kind corresponds to the total magnetization
reversal accompanied by the broad band superradiance. It is
described by a rigorous model \cite{CG-SR} which is reviewed in
Section II. In essence, if one neglects the electromagnetic
radiation, the crossing of the ($m,m'$) resonance by the magnetic
field sweep, Fig.\ \ref{fig:spl}, is described by the Landau-Zener
theory \cite{LZ}. When the coupling between the spins and the
electromagnetic radiation is taken into account, the magnetic
state resulting from the LZ transition relaxes towards the lowest
energy state via superradiance. The rate of the superradiant
decay, as well as the time dependence of the relaxation, are
sensitive to the parameters of the crystal and to the shape of the
magnetic field pulse. Our goal is to compute the time dependence
of the radiation power and the total radiated energy as functions
of the field-sweep rate, the tunnel splitting, and the size of the
crystal. This is done in Section III by analytical and numerical
methods. Practical implications of our findings are discussed in
Section IV.

\section{Collective Landau-Zener relaxation}

Consider a crystal of $N$ magnetic molecules occupying an $m$
magnetic state that is close to the resonance with the $m'$ state,
e.g. $m=-S, m'=S$ in Fig.\ \ref{fig:levels}. We shall assume that
the molecules weakly interact with each other through the
electromagnetic field. As has been shown in Ref. \cite{CG-SR}, the
quantum magnetic relaxation of such a crystal satisfies the
Landau-Lifshitz equation:
\begin{equation}\label{LL}
\dot{\bf n}=\gamma [{\bf n\times H}_{\rm eff}] - \alpha\gamma[{\bf
n}{\times}[{\bf n}{\times}{\bf H}_{\rm eff}]] \;.
\end{equation}
Here ${\bf n}$ is a unit vector of the pseudospin describing the
two-state system, such that $n_z = -1$ corresponds to all
molecules in the $m$-state, while $n_z = 1$ corresponds to all
molecules in the $m'$-state, $\gamma = g {\mu}_{B}/\hbar$ is the
gyromagnetic ratio, the effective magnetic field is given by
\begin{equation}\label{HeffDef}
g\mu_B {\bf H}_{\rm eff} = \Delta {\bf e}_x+W{\bf e}_z \;,
\end{equation}
with $\dot{W}(t) = \frac{1}{2} g {\mu}_B (m' - m) {\dot{H}}_z$
being the energy sweep rate (see Fig.\ \ref{fig:spl}), and $\alpha
\ll 1$ is a dimensionless effective damping coefficient,
\begin{equation}\label{alphaRes}
\alpha=\frac{1}{24}N (m'-m)^{2} g^2 {\alpha}_o
\left(\frac{\Delta}{m_e c^2}\right)^2 \;,
\end{equation}
with ${\alpha}_o = {e^2}/{\hbar c} \approx 1/137$ being the fine
structure constant. Note that $\alpha$ is independent of the
magnetic field.

The first term in Eq.\ (\ref{LL}) gives dissipationless
Landau-Zener transitions when the field is swept through the
resonance such that $W=W(t)$ satisfies $W(\pm\infty)=\pm\infty$,
and the initial condition is ${\bf n}(-\infty)=-{\bf e}_z$.
Indeed, at $\alpha = 0$ the Schr\"odinger equation for a two-level
system is equivalent to the equation for a precessing spin. The
probability $p(t)$ for the molecule to stay in the initial state
is given by
\begin{equation}\label{Pvianz}
p(t)=[1-n_z(t)]/2 \;.
\end{equation}
For $W(t)=vt$, one obtains the famous Landau-Zener result
\cite{LZ}: $p(\infty)\equiv p_{LZ}=\exp(-\epsilon)$, where
\begin{equation}\label{epsilon}
\epsilon = \frac{\pi \Delta^2}{2\hbar v} \;.
\end{equation}
\begin{figure}
\unitlength1cm
\begin{picture}(12,4.9)
\centerline{\psfig{file=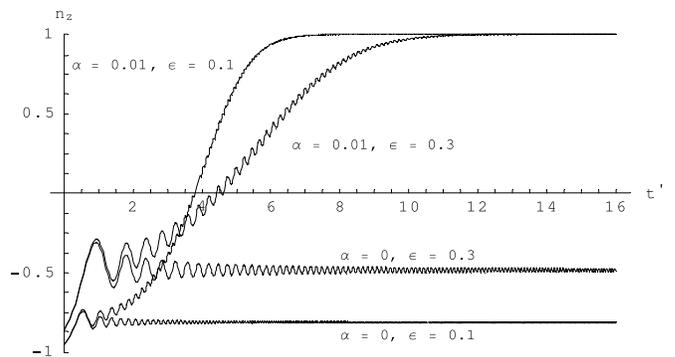,angle=0,width=8.8cm}}
\end{picture}
\caption{\label{fig:nzgraph} Time dependence of the magnetization
reversal for two values of $\epsilon$ due to pure Landau-Zener
relaxation of individual molecules ($\alpha = 0$) and due to
collective relaxation via superradiance ($\alpha = 0.01$).}
\end{figure}
The Landau-Zener effect corresponds to only partial magnetization
reversal,
\begin{equation}\label{nLZ}
n_{z}^{LZ}(\infty) = 1 - 2\exp(-\epsilon)\;,
\end{equation}
see Fig.\ \ref{fig:nzgraph} at $\alpha = 0$. The value of
$n_{z}^{LZ}(\infty)$ is close to $-1$ at $\epsilon \ll 1$, that is
for the fast field sweep. In this case most of the molecules,
after crossing the resonance, remain in the initial $m$-state by
passing from the lower to the upper branch in Fig.\ \ref{fig:spl}.
On the contrary, for a slow sweep, that is when $\epsilon \gg 1$,
most of the molecules follow the lower branch in Fig.\
\ref{fig:spl} and the final state of the crystal is exponentially
close to $n_{z}^{LZ}(\infty) = 1$.

The second term in Eq.\ (\ref{LL}) describes collective magnetic
relaxation via Dicke superradiance. Due to this term the
magnetization of the entire crystal at long times reverses
completely  to $n_z(\infty) = 1$, as is shown in Fig.\
\ref{fig:nzgraph}, for a finite $\alpha$. The collective
relaxation due to superradiance is significant for $ \epsilon
\lesssim 1$, that is, when $n_{z}^{LZ}(\infty)$ is not very close
to $1$. Thus, the observation of the superradiance requires a fast
field sweep.

\section{Radiation power}

The power of the superradiance described by Eq.\ (\ref{LL}) can be
obtained from the classical formula for the magnetic dipole
radiation \cite{CG-SR}:
\begin{equation}\label{MDP}
P(t) = [2/(3c^3)]\ddot m_z^2(t)\;,
\end{equation}
where
\begin{equation}\label{m_z}
m_z(t) =\frac{1}{2}N(m-m')g\mu_B n_z(t) \,.
\end{equation}
Close to the ($m,m'$) resonance, nearly any field sweep of
practical interest is linear in time, $W=vt$. It is convenient to
use dimensionless variables:
\begin{equation}\label{dimensionless}
t' = \frac{t \Delta }{ \hbar}\;, \;\;\;\;\;W'(t') =  \frac{v t}
{\Delta} = \frac{\hbar v t'}{{\Delta}^2} = \frac{\pi t'}{ 2
\epsilon} \;.
\end{equation}
In terms of these variables Eq.\ (\ref{LL}) and Eq.\ (\ref{MDP})
become
\begin{equation}\label{LL-dimensionless}
\frac{d{\bf n}}{dt'} = [{\bf n} \times ({\bf e}_x + W'(t'){\bf
e}_z)] - \alpha[{\bf n}{\times}[{\bf n}{\times}({\bf e}_x +
W'(t'){\bf e}_z)]]\;
\end{equation}
and
\begin{equation}\label{power}
P = {\alpha} N {\hbar}^{-1}{\Delta}^{2}
\left(\frac{d^2n_z}{d{t'}^{2}}\right)^2\;.
\end{equation}
The total emitted energy, $E = \int dt \, P(t)$, is given by
\begin{equation}\label{E}
E = \alpha N \Delta \, E'\;,
\end{equation}
where we have introduced dimensionless
\begin{equation}\label{E'}
E' = \int dt'\left(\frac{d^2n_z}{d{t'}^{2}}\right)^2 \;.
\end{equation}

\subsection{Analytical}

\begin{figure}
\unitlength1cm
\begin{picture}(8.4,6.25)
\centerline{\psfig{file=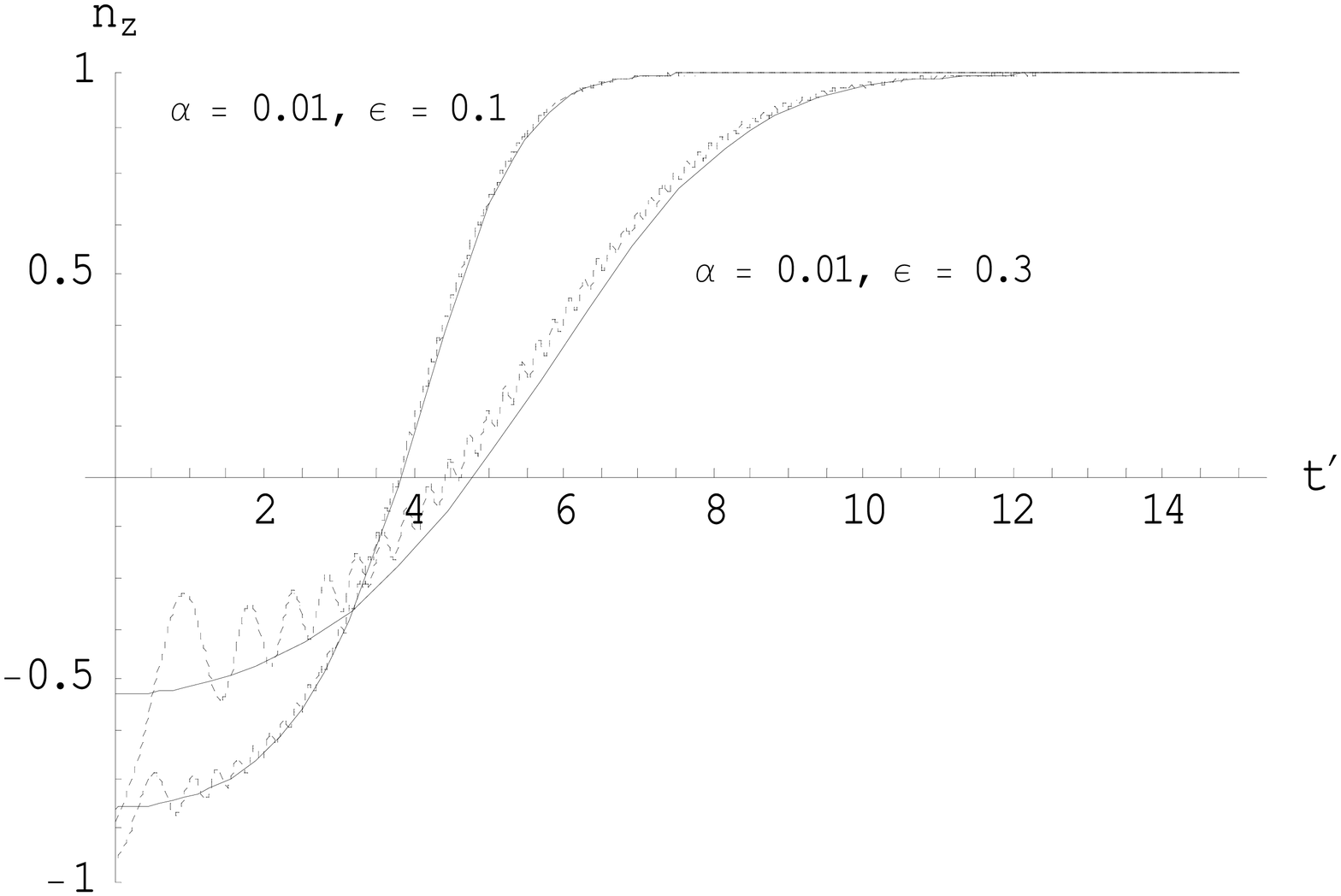,angle=0,width=8.7cm}}
\end{picture}
\caption{\label{fig:nzgraph1} Approximate analytical solution for
$n_z(t')$ averaged over oscillations, Eq.\ (\ref{nz-solution}),
for two values of $\epsilon$ and $\alpha = 0.01$ (solid line). The
numerical solution of Eq.\ (\ref{LL-dimensionless}) is shown by
the dash line.}
\end{figure}

We shall start by developing an analytical approximation for the
practical case of $\epsilon < 1$ and $\alpha \ll 1$. The time
interval of interest is the one past the Landau-Zener transition:
$W' \gg 1$. In this case, retaining the leading terms in Eq.\
(\ref{LL-dimensionless}), we get
\begin{eqnarray}\label{system}
\frac{dn_x}{dt'} & = & W'n_y \label{x} \\
\frac{dn_y}{dt'} & = & -W'n_x \label{y} \\
\frac{dn_z}{dt'} & = & - n_y + {\alpha} W'(t')(1 - n_z^2) \;.
\label{z}
\end{eqnarray}
These equations show that $n_x$ and $n_y$ oscillate rapidly in
time, while $n_z$, in accordance with Fig.\ \ref{fig:nzgraph}, has
a slowly varying average. Averaging Eq.\ (\ref{z}) over the period
of oscillations of $n_y$, one obtains:
\begin{equation}\label{z-average}
\frac{d{\bar{n}}_z}{dt'} = {\alpha} W'(t')(1 - {\bar{n}}_z^2) \;.
\end{equation}
Eq.\ (\ref{z-average}) describes the superradiant stage of the
evolution of ${\bar{n}}_z$. Therefore, it must be solved with the
initial condition ${\bar{n}}_z = n_{z}^{LZ}$ at $t=0$. At small
$\epsilon$, Eq.\ (\ref{nLZ}) gives for that initial condition:
\begin{equation}\label{initial}
{\bar{n}}_z(0) = -1 +2\epsilon\;.
\end{equation}
The corresponding solution of Eq.\ (\ref{z-average}) reads
\cite{Leuven}
\begin{equation}\label{nz-solution}
{\bar{n}}_z(t') = \tanh \left(\frac{\alpha \pi {t'}^2}{4 \epsilon}
- \frac{1}{2}\ln\frac{1}{\epsilon} \right) \;.
\end{equation}
It is shown by the solid line in Fig.\ \ref{fig:nzgraph1}. For
$\epsilon = 0.1$ Eq.\ (\ref{z-average}) is, clearly, a good
approximation to the full solution averaged over oscillations. As
$\epsilon$ increases, some discrepancy is observed. One can
improve the analytical approximation by writing $n_z = {\bar{n}}_z
+ \delta n_z$ and solving Eq.\ (\ref{z}) through iterations, but
this, at the end, will require a numerical integration, so that
the improvement obtained by this method does not give much
advantage over the direct numerical solution of Eq.\
(\ref{LL-dimensionless}).

The approximate solution of Eqs.\ (\ref{x}) and (\ref{y}),
satisfying ${\bf n}^2 =1$, is given by
\begin{eqnarray}\label{xy}
n_x & = & \sqrt{1-{\bar{n}}_z^2} \sin \left(\frac{\pi {t'}^2}{4\epsilon} + \phi_0 \right) \nonumber \\
n_y & = & -\sqrt{1-{\bar{n}}_z^2} \cos \left(\frac{\pi
{t'}^2}{4\epsilon} + \phi_0 \right) \;,
\end{eqnarray}
where $\phi_0$ is a phase which we are not attempting to compute
analytically. One can see from Eqs.\ (\ref{nz-solution}) and
(\ref{xy}) that ${\bar{n}}_z$, indeed, changes slowly with time,
compared to the oscillations of $n_x(t')$ and $n_y(t')$, because
of the condition $\alpha \ll 1$.

Let us now turn to the analytical approximation for the power and
the total radiated energy. It is easy to see from Eqs.\
(\ref{z})-(\ref{xy}) that the main contribution to
$d^2n_z/d{t'}^2$ is determined by the rapidly oscillating
$n_y$-term in the right-hand-side of Eq.\ (\ref{z}):
\begin{equation}\label{second-derivative}
\frac{d^2n_z}{d{t'}^2} = - \left( \frac{\pi t'}{ 2 \epsilon}
\right)\frac{\sin[({\pi {t'}^2}/{4\epsilon}) + \phi_0]}{\cosh
[({\alpha \pi {t'}^2}/{4 \epsilon}) + \frac{1}{2}\ln{\epsilon}]}
\;.
\end{equation}
Substituting this expression into Eq.\ (\ref{E'}) and replacing
the rapidly oscillating ${\sin}^2(\pi {t'}^2/{4\epsilon})$ under
the integral by $1/2$, one finally obtains:
\begin{equation}\label{E'-computed}
E' =
\frac{\sqrt{\pi}}{{\epsilon}^{1/2}{\alpha}^{3/2}}\int_0^{\infty}
\frac{x^2\;dx}{{\cosh}^2 (x^2 + \frac{1}{2}\ln{\epsilon})} \;.
\end{equation}
Equations (\ref{power}), (\ref{E}), (\ref{E'}),
(\ref{second-derivative}), and (\ref{E'-computed}) give the
dependence of the radiation power and the total emitted energy on
the field sweep rate and on the parameters of the crystal.

\subsection{Numerical}

We shall now compute $P(t)$ and $E$ by numerical integration of
Eq.\ (\ref{LL-dimensionless}), and compare them with our
analytical findings.

\begin{figure}
\unitlength1cm
\begin{picture}(11,6)
\centerline{\psfig{file=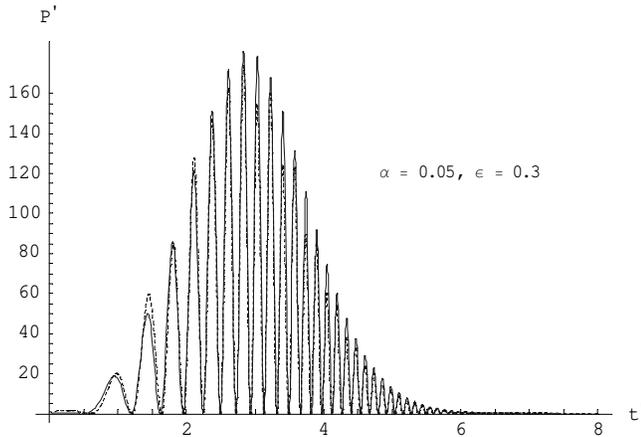,angle=0,width=8.8cm}}
\end{picture}
\caption{\label{fig:powergraph} Time dependence of the reduced
radiation power, $P' = (d^2n_z/d{t'}^2)^2$ at $\epsilon = 0.3$ and
$\alpha = 0.05$. Solid line represents numerical results. Dash
line corresponds to Eq.\ (\ref{second-derivative}) at ${\phi}_0 =
2.576$.}
\end{figure}

The time dependence of the reduced power, $P' =
(d^2n_z/d{t'}^2)^2$, is shown in Fig.\ \ref{fig:powergraph}.
The comparison with Eq.\ (\ref{second-derivative}) is performed by
fitting the value of ${\phi}_0$ until the match with the numerical
solution of Eq.\ (\ref{LL-dimensionless}) for $(d^2n_z/d{t'}^2)^2$
is obtained. Even for $\epsilon$ as large as $0.3$ the agreement
of the numerical results with the analytical formula is rather
good. Note that the oscillation of the power in time is a quantum
effect related to the oscillation of $n_z$.

\begin{figure}
\unitlength1cm
\begin{picture}(11,6)
\centerline{\psfig{file=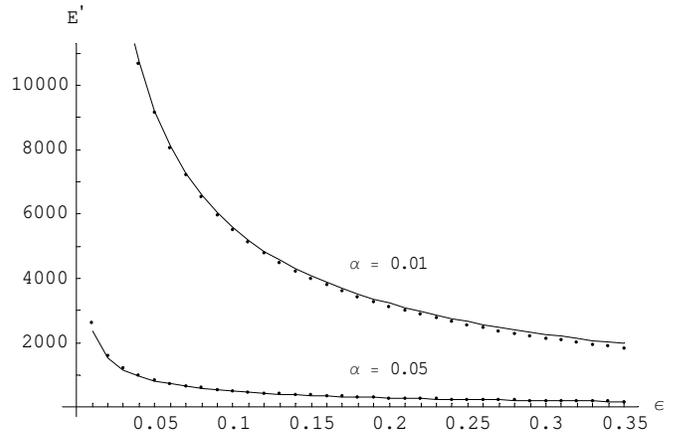,angle=0,width=8.8cm}}
\end{picture}
\caption{\label{fig:epsilongraph} The $\epsilon$ dependence of the
total emitted energy at two values of $\alpha$. Points represent
numerical results. Solid line corresponds to Eq.\
(\ref{E'-computed}).}
\end{figure}

\begin{figure}
\unitlength1cm
\begin{picture}(11,6.7)
\centerline{\psfig{file=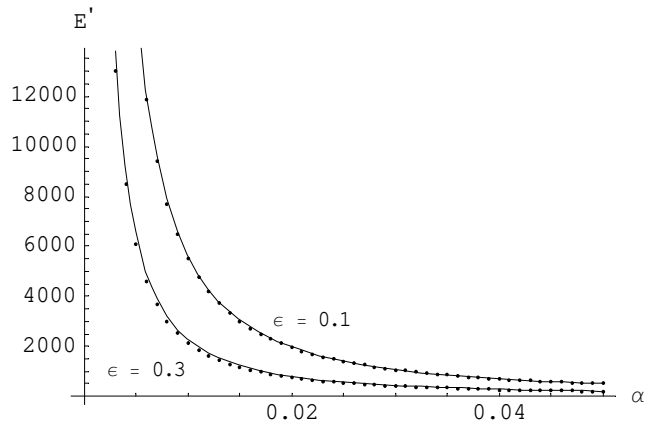,angle=0,width=8.6cm}}
\end{picture}
\caption{\label{fig:alphagraph} The $\alpha$ dependence of $E'$ at
two values of $\epsilon$. Points represent numerical results.
Solid line corresponds to Eq.\ (\ref{E'-computed}).}
\end{figure}

Fig.\ \ref{fig:epsilongraph} shows the dependence of the total
emitted energy on the parameter $\epsilon$, that is, on the
inverse field sweep rate.
Fig.\ \ref{fig:alphagraph} shows the dependence of $E'$ on the
parameter $\alpha$.

\begin{figure}
\unitlength1cm
\begin{picture}(8.5,6.2)
\centerline{\psfig{file=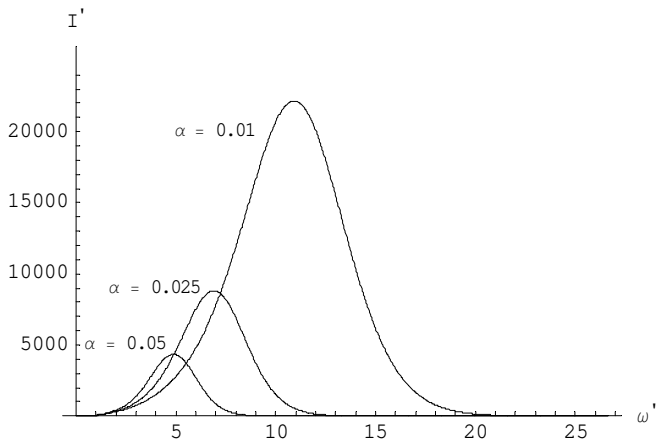,angle=0,width=8.7cm}}
\end{picture}
\caption{\label{fig:spectral} Spectral function $I'({\omega}')$
for three values of alpha at $\epsilon = 0.1$. }
\end{figure}

\begin{figure}
\unitlength1cm
\begin{picture}(11,6.5)
\centerline{\psfig{file=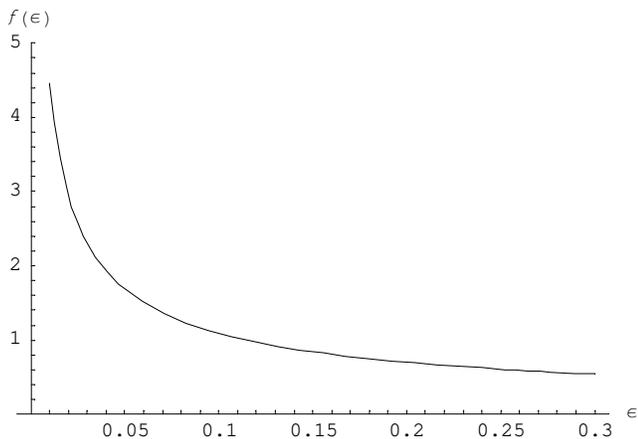,angle=0,width=9cm}}
\end{picture}
\caption{\label{fig:f} Dependence of $f$ of Eq.\
(\ref{frequency-exact}) on $\epsilon$. }
\end{figure}

The question of significant importance for experiment is the
spectral composition of the radiation. The total emitted energy
can be presented as
\begin{equation}\label{E-spectral}
E = \int d{{\omega}} I({\omega}) \;,
\end{equation}
where
\begin{equation}\label{I}
I(\omega) = {\hbar}\, {\alpha} N I'({\omega}')
\end{equation}
is the spectral power. Here $I'({\omega}')$ is a dimensionless
function of the dimensionless frequency, ${\omega}' = {\hbar
\omega}/\Delta$. It must be computed via the Fourier transform of
$d^2n_z/d{t'}^2$:
\begin{equation}\label{I'}
I'({\omega}') = \frac{1}{2\pi} \left|\int dt' e^{i {\omega}' t'}
\left(\frac{d^2 n_z}{d {t'}^2}\right)\right|^2 \;.
\end{equation}
This function is shown in Fig.\ \ref{fig:spectral}.
The peak of the power occurs at
\begin{equation}\label{frequency-exact}
{\hbar}{\omega}_{max} = \frac{\Delta}{\sqrt{\alpha}}f(\epsilon)\;.
\end{equation}
This scaling of ${\hbar}{\omega}_{max}$ on $\alpha$ follows from
Eq.\ (\ref{LL-dimensionless}). The function $f(\epsilon)$,
computed numerically, is shown in Fig.\ \ref{fig:f}.

\section{Discussion}

The formulas and the numerical results obtained above are valid if
the energy distance between the resonant levels, $W$, is small
compared to the distance between the adjacent $m$-levels. For
numerical estimates, we shall stick to the $(-S,S)$-resonance. The
conclusions of this section, however, will apply to other
resonances as well. For the model illustrated by Eq.\
(\ref{biaxial}) and Fig.\ \ref{fig:levels}, the distance between
the $m = -S$ level and the $m = -S + 1$ level is $(2S-1)D \approx
2SD$. The validity condition we are looking for is then $W(t) \ll
2SD$. For $W = vt$ one should verify this condition at $W_{max} =
vt_{max}$, where $t_{max} = {\hbar}{t'}_{max}/\Delta$ is the time
when the superradiance drops exponentially due to the hyperbolic
cosine in Eq.\ (\ref{second-derivative}). According to this
equation and Eq.\ (\ref{dimensionless}), ${t'}_{max} \sim
\sqrt{\epsilon/\alpha}$ and $W_{max} \sim {\Delta}/\sqrt{\epsilon
\alpha}$. Substituting this into $W_{max} \ll 2SD$ and using Eq.\
(\ref{alphaRes}) for ${\alpha}$, one obtains the validity
condition in the form of the lower bound on the total number of
molecules:
\begin{equation}\label{validity}
N \gg \frac{1}{\epsilon {\alpha}_o}\left(\frac{m_e
c^2}{U_a}\right)^2 \;,
\end{equation}
where $U_a = DS^2$ is the energy barrier between $m = \pm S$
states due to magnetic anisotropy. Eq.\ (\ref{validity}) shows
that a high magnetic anisotropy and a not very small ${\epsilon}$
are needed if the size of the system is to remain within
reasonable limits. The optimal would be $\epsilon \sim 1$ since,
according to Eq.\ (\ref{nLZ}), $\epsilon \lesssim 1$ (that is, a
sufficiently high field-sweep rate) is needed to create an inverse
population of spin levels. For Mn$_{12}$ and Fe$_8$, the
anisotropy barrier is of order $60$K and $30$K respectively, and
the lower bound on $N$, according to Eq.\ (\ref{validity}), must
be between $10^{18}$ and $10^{19}$ molecules. With account of the
unit cell volume ($3.7\,$nm$^3$ and $2.0\,$nm$^3$ for Mn$_{12}$
and Fe$_8$, respectively) this translates into a volume of order
or greater than $1$mm$^3$. Remarkably, this agrees with the
reported lower bound on the volume of the crystal (or crystal
assembly) that shows evidence of electromagnetic radiation during
magnetization reversal \cite{Tejada1,Tejada2,Leuven}.

We shall now estimate the total emitted energy and the power of
the radiation. According to Eq.\ (\ref{E'-computed}), $E' \sim
{\epsilon}^{-1/2}{\alpha}^{-3/2}$. This gives for $E$ of Eq.\
(\ref{E}): $E \sim N {\Delta}/\sqrt{\epsilon \alpha}$. With the
help of Eq.\ (\ref{alphaRes}) one obtains:
\begin{equation}\label{E-estimate}
E \sim {\epsilon}^{-1/2} N^{1/2} m_e c^2 \;.
\end{equation}
For the purpose of the order-of-magnitude estimate we have dropped
the factor $g S \sqrt{{\alpha}_0}$ of order unity. Notice that the
total emitted energy is proportional to the square root of the
crystal volume. At ${\epsilon} \sim 1$ and $N \sim 10^{18}$, Eq.\
(\ref{E-estimate}) provides $E \sim 0.1\,$mJ.

According to Eqs.\ (\ref{power}) and (\ref{second-derivative})
(see also Fig.\ \ref{fig:powergraph}) the power of the radiation
oscillates in time. In most cases, observation of these
oscillations must be impeded by the finite time resolution of the
measuring equipment, so that only the envelope of the curve shown
in Fig.\ \ref{fig:powergraph} will be observed. The peak power can
be estimated as $P_{max} \sim E/t_{max} \sim N {\Delta}^{2}/\hbar
\epsilon$. Substituting here ${\epsilon}$ of Eq.\ (\ref{epsilon}),
one obtains
\begin{equation}\label{P-estimate}
P_{max} \sim N v \sim {\delta}M \frac{dH}{dt} \;,
\end{equation}
where we have introduced ${\delta}M = g {\mu}_B (m' - m) N$, the
change in the total magnetic moment of the crystal due to
collective electromagnetic relaxation. Note that the relations $E
\propto \sqrt{N}$ and $P_{max} \propto N$ are specific to the
radiation problem we have studied. For an assembly of a few
mm-size crystals Eq.\ (\ref{P-estimate}) gives $P_{max} \sim
10\,{\mu}$W at a typical laboratory field sweep rate of
$0.01\,$T/s \cite{Tejada1,Tejada2} and $P_{max} \sim 1\,$W for a
fast field pulse, $dH/dt \sim 10^3$T/s \cite{Leuven}. Note,
however, that in the case of an ultrafast sweep the condition
${\epsilon} \sim 1$ can be satisfied only by a large tunnel
splitting ${\Delta}$, making the preparation of the initially
magnetized state less simple than in the case of small $\Delta$.

During the adiabatic sweep, the frequency of the radiation is
determined by the distance between the spin levels, $\hbar{\omega}
= \sqrt{{\Delta}^2 + W^2}$. For the most of the relaxation
process, $W \gg {\Delta}$, and, thus, $\omega = W(t)/\hbar$. The
peak of the spectral power, Fig.\ \ref{fig:spectral}, corresponds
to ${\hbar}{\omega}_{max} \sim W(t_{max}) \sim \Delta/ \sqrt{
\epsilon \alpha}$. Up to a factor of order unity, that depends
logarithmically on $\epsilon$, this coincides with Eq.\
(\ref{frequency-exact}). The logarithmic difference of
$f(\epsilon)$ in Fig.\ \ref{fig:f} from $1/\sqrt{\epsilon}$ is due
to $\ln{\epsilon}$ in Eq.\ (\ref{second-derivative}). With the
help of Eq.\ (\ref{alphaRes}), we obtain that by order of
magnitude
\begin{equation}\label{frequency}
{\omega}_{max} \sim \frac{m_ec^2}{\hbar \sqrt{\epsilon N}} \;,
\end{equation}
where we again omitted the factor $g S \sqrt{{\alpha}_0}$ of order
unity. For ${\epsilon} \sim 1$ and $N \sim 10^{18}$, required to
produce significant radiation (see above), this frequency is in
the terahertz range.

For the radiation to be coherent, the inhomogeneous broadening of
$\epsilon$ must be small throughout the crystal. This translates
into narrow distribution of the tunnel splitting and narrow
distribution of the magnetic field felt by the spins. Both
conditions must be satisfied in Fe$_8$ because they are also the
necessary conditions for the Berry phase effect observed in that
system \cite{WS}. In Mn$_{12}$ the situation is less clear due to
solvent disorder, large hyperfine interactions, dislocations,
etc., which result in a distribution of ${\Delta}$
\cite{dislocations,Myriam,Cornia,Hill-Mn,Kent-Mn}. The suitability
of Mn$_{12}$ for the study of superradiance depends on whether the
distribution of ${\Delta}$ is continuous or consists of a finite
number of narrow lines due to, e.g., finite number of nuclear spin
states, presence of isomers in the structure of the molecule
\cite{Cornia}, etc. Systems with narrow distribution of $\Delta$
probably exist among hundreds of new molecular magnets synthesized
in recent years. In such systems, narrow distribution of the
magnetic field should be achieved automatically when the spins of
the initially saturated sample rotate coherently due to
superradiance.\\

\section{Conclusions}

We have studied magnetic relaxation via coherent electromagnetic
radiation, produced by the magnetic field sweep in a crystal of
molecular magnets on crossing the tunneling resonance. We find
that the effect exists starting roughly with crystals (or crystal
assembly) of millimeter size. The radiation is broadband with the
cutoff in the terahertz range. The power of the radiation is
proportional to the field-sweep rate and ranges from microwatts to
watts for the existing sweep rates.

\section{Acknowledgments}

This work has been supported by the NSF Grant No. EIA-0310517.

\end{document}